\documentclass[twocolumn]{aastex61}

\usepackage[version=4]{mhchem}
\usepackage{natbib}
\usepackage{amsmath}
\usepackage{enumerate}
\bibpunct{(}{)}{;}{a}{}{,}

\def\arcsec{\hbox{$^{\prime\prime}$}}

\shorttitle{Late-Time Photometry of SN\,2013aa}
\shortauthors{Jacobson-Galan et al.}

\begin{document}

\title{Constraining Type Ia Supernova Progenitor Scenarios with Extremely Late-time Photometry of Supernova SN\,2013\MakeLowercase{aa}}

\begin{abstract}
  We present \textit{Hubble Space Telescope} observations and
  photometric measurements of the Type Ia supernova (SN~Ia) SN~2013aa
  1500~days after explosion. At this epoch, the luminosity is
  primarily dictated by the amounts of radioactive ${}^{57}\textrm{Co}$ and
  ${}^{55}\textrm{Fe}$, while at earlier epochs, the luminosity depends on the amount of radioactive ${}^{56}\textrm{Co}$. The ratio of odd-numbered to even-numbered isotopes depends significantly on the
  density of the progenitor white dwarf during the SN explosion, which, in turn, depends on the details of the progenitor system at the time of ignition.  From a comprehensive analysis of the entire light curve of SN~2013aa, we measure a
  $M({}^{57}\textrm{Co})/M({}^{56}\textrm{Co})$ ratio of
  $0.02^{+0.01}_{-0.02}$, which indicates a relatively low central
  density for the progenitor white dwarf at the time of explosion,
  consistent with double-degenerate progenitor channels. We estimate
  $M({}^{56}\textrm{Ni}) = 0.732 \pm 0.151\:\mathrm{M_{\odot}}$, and
  place an upper limit on the abundance of ${}^{55}\textrm{Fe}$.  A
  recent study reported a possible correlation between
  $M({}^{57}\textrm{Co})/M({}^{56}\textrm{Co})$ and
  stretch for four SNe~Ia.  SN\,2013aa, however, does
  not fit this trend, indicating either SN\,2013aa is an
  extreme outlier or the correlation does not hold up with a larger
  sample.  The $M({}^{57}\textrm{Co})/M({}^{56}\textrm{Co})$ measured for the expanded sample of SNe~Ia with photometry at extremely late times has a much larger range than that of explosion models, perhaps limiting conclusions about SN~Ia progenitors drawn from extremely late-time photometry.
 
\end{abstract}


\correspondingauthor{Wynn Jacobson-Galan}
\email{wjacobso@ucsc.edu}

\author{Wynn V. Jacobson-Gal\'an}
\affil{Department of Astronomy and Astrophysics, University of California, Santa Cruz, CA 95064,
USA}

\author{Georgios Dimitriadis}
\affil{Department of Astronomy and Astrophysics, University of California, Santa Cruz, CA 95064,
USA}

\author{Ryan J. Foley}
\affil{Department of Astronomy and Astrophysics, University of California, Santa Cruz, CA 95064,
USA}

\author{Charles D. Kilpatrick}
\affil{Department of Astronomy and Astrophysics, University of California, Santa Cruz, CA 95064,
USA}

\section{Introduction} \label{sec:intro}

Type Ia Supernova (SNe~Ia) are the result of a thermonuclear explosion
of a carbon-oxygen white dwarf (WD) in a binary system \citep{Hoyle60,
  Colgate69, Woosley86}.  While the applications of SNe Ia as
standardizable candles are far reaching in the realm of cosmology
\citep[e.g.,][]{reiss98, perlmutter99}, the exact nature of the
explosion and the progenitor system, and in particular the binary
companion, are still an open question \citep[see e.g.,][]{maoz14}.

There are several ways to potentially produce a SN~Ia \citep{woosley-taam86}.  In
these models, the (primary) WD is usually either very close to the
Chandrasekhar mass, having undergone a simmering stage \citep{piro308,piro08} and having a
high central density, or below the Chandrasekhar mass with a lower
central density \citep{iben84, woosley04}.  The details of the explosive nuclear
burning depends critically on the central density.  In particular,
explosions with higher central densities will produce more Fe-group
elements with an odd number of nucleons \citep{iwamoto99,seitenzahl13a}. While other aspects of the explosion have a larger effect on the amount of odd-numbered, radioactive isotopes produced (e.g., metallicity or M($^{56}$Ni)), measuring the mass of these isotopes
can distinguish explosion models.

In addition to different explosions, there are fundamentally different
progenitor channels for SNe~Ia.  The single-degenerate (SD) and
double-degenerate (DD) channels, which have non-degenerate and WD
companion stars, respectively.  The DD channel will naturally have a
sub-Chandrasekhar mass primary and a relatively low central density \citep{pakmor10,pakmor11}.
While some SD systems might result in a sub-Chandrasekhar mass
explosion, the classical model involves a Chandrasekhar-mass WD and a
high central density \citep{khokhlov00,han04}.

The single-degenerate (SD) model argues that the explosion is triggered
by a high central density, delayed detonation of a
near-Chandrasekhar-mass WD as it accretes material and energy from
main-sequence or larger star (\citealt{whelan73}, \citealt{khokhlov91}). Alternatively, the double-degenerate (DD) model
consists of a low central density, violent merger of two,
sub-Chandrasekhar-mass WD stars (\citealt{webbink84}, \citealt{pakmor12}). While both are accepted theoretical predictions,
the direct detection of the progenitor system is difficult, with most
DD models leaving no post-explosion indication of the system
responsible. There have, however, been recent constraints placed on
the direct detection of progenitor systems following SD models
(\citealt{chomiuk16}, \citealt{maguire16}). Fortunately, other methods
of progenitor system constraint come from the unique modeling of these
explosions by \citet{ropke12} and \citet{seitenzahl13}, all of which are
verifiable via the study of radioactive decay in late-time bolometric
light curves of SNe Ia.

By Arnett's Law, the bolometric luminosity produced at peak magnitude is proportional to the rate of energy deposition by the radioactive decay chain $\ce{{}^56Ni ->[ t_{1/2}=6.08 \textrm{d} ] {}^56Co ->[ t_{1/2}=77.2 \textrm{d} ] {}^56Fe }$ \citep{arnett82}. While Arnett's Law is an approximation, the decay of ${}^{56}\textrm{Ni}$ remains the most prominent source of heating in SNe Ia and produces primarily $\gamma$-rays and positrons, whose energies are deposited and thermalized in the expanding ejecta \citep{Seitenzahl17arXiv}. Not only can the total mass of ${}^{56}\textrm{Ni}$ be determined from the peak luminosity, the isotopic yields generated in decay chains ${}^{57}\textrm{Co} \rightarrow {}^{57}\textrm{Fe}$ and ${}^{55}\textrm{Fe} \rightarrow {}^{55}\textrm{Mn}$ can be indirectly detected from the light curve evolution of SNe Ia at epochs $>300$ days after explosion \citep{seitenzahl2009}. Model analysis has shown that the mass ratios of these nucleosyntheic yields differ between single and double degenerate explosion models, thus making them extremely useful in identifying the pre-explosion SNe Ia progenitor systems \citep{ropke12}. 

Testing each model requires precise photometric data from continuous observations of nearby SNe Ia $>400$ days after peak luminosity. This is a challenging effort due to the variability of SNe Ia explosions coupled with the ability to perform accurate photometric measurements at late enough epochs to detect the radioactive decay of isotopes other than ${}^{56}\textrm{Ni}$. Nonetheless, a few significant studies have been recently performed on SNe Ia in close proximately to us and with multiple broad band photometric detections produced at late epochs. 

SN\,2011fe remains to be one of the most highly studied late-time SNe Ia, with numerous examinations of radioactive decay channels since its nearby discovery \citep{kasen13}. \citet{shappee17} were able to detect abundances of ${}^{56}\textrm{Co}$ and ${}^{57}\textrm{Co}$ as well as place an upper limit on the mass of ${}^{55}\textrm{Fe}$, while indicating that the fits to the data preferred a DD explosion model. A similar study by \citet{dimitriadis17} examined the near infrared contribution to the bolometric luminosity of SN\,2011fe, but found a contradicting alignment to the single-degenerate explosion model of a high central density white dwarf star. 

Further examinations of extremely late-time supernovae also make predictions of the pre-explosion progenitor system. \citet{Graur16} finds a distinct detection of ${}^{57}\textrm{Co}$ in the light curve of SN\,2012cg and predicts a single-degenerate explosion mechanism. The analysis of SN\,2014J makes similar conclusions in their determination of mass ratios that prefer a high central density explosion model \citep{yang17}. Alternatively, the mass ratio found in SN\,2015F by \citet{graur17} shows evidence for a double degenerate merger of two white dwarfs. \citet{graur17} also examines the relationship between the calculated light curve stretch and ${}^{57}\textrm{Co}$/${}^{56}\textrm{Co}$ in all four late-time SNe Ia, the implications of which we will discuss as it relates to SN\,2013aa. 

The detection of SN\,2013aa at a phase of $\sim1500$ days presents a unique opportunity to examine the nucleosynthetic yields of late-time decay. SN\,2013aa is the fifth SNe Ia to be observed at an epoch $>1000$ days, with a photometric detection at the second latest phase next to SN\,2011fe. The measured late-time bolometric luminosity, combined with early-time data, allows for a fitted calculation of radionuclide abundances powering the light curve. The mass ratios of ${}^{56}\textrm{Co}$, ${}^{57}\textrm{Co}$, and ${}^{55}\textrm{Fe}$ found in SN\,2013aa can then compared with explosion models as an indicator of the progenitor system. With only four recorded late-time SNe Ia prior to SN\,2013aa, this analysis will contribute to the understanding of late-time trends in the light curves of SNe Ia.

In Section \ref{sec:observations} we present observations and data reduction of SN\,2013aa. In Section \ref{sec:Analysis} we discuss the calculation of radioactive isotope abundances. In Section \ref{sec:Discussion} we examine the implications of measured mass ratios in the context of explosion models and other late-time studies.

\section{Observations} \label{sec:observations}

In this section, we briefly introduce SN\,2013aa, presenting the
published photometric and spectroscopic data and basic parameters from
early-time data. We also present late-time \textit{HST} photometry.

\subsection{Early-time data (up to 400~days)}
\label{subsec:early_data}

SN\,2013aa was discovered by the Backyard Observatory Supernova Survey
(BOSS) on 2013 February 13 \citep{parker13} and confirmed to be a
SN~Ia on 2013 February 16 \citep{parrent13}. SN\,2013aa is located in
the barred spiral galaxy NGC\,5643, 74\arcsec\:West and
180\arcsec\:South from the galactic center \citep{Graham17arXiv}.  Another SN~Ia,
SN\,2017cbv, is in the same galaxy, providing an independent distance
estimate to SN\,2013aa (Shappee et~al., in preparation).  Applying the SALT2 algorithm \citep{Guy07} to the SN\,2017cbv data, we determine that the
distance to NGC~5643 is $13.95 \pm 0.35$~Mpc, corresponding to a
distance modulus of $\mu = 30.72 \pm 0.05$~mag.  Primary parameters of
SN\,2013aa and its host galaxy, NGC\,5643, are reported in
Table~\ref{tbl:params}.

\begin{table}[ht]
\begin{center}
\caption{Main Parameters of SN\,2013aa and Host Galaxy \label{tbl:params}}
\vskip0.1in
\begin{tabular}{lccccccccc}
\hline
\hline
Host Galaxy &  &  & &  & & & &  &  NGC 5643 \\ 
Galaxy Type &  &  & &  & & & &  &  SAB(rs)c \\  
Redshift &  &  & &  & & & &  &  $0.003999 \pm 0.000007$\\  
Distance &  &  & &  & & & &  &  $13.95 \pm 0.3$~Mpc\\ 
Distance Modulus, $\mu$ &  &  & &  & & & &  &  $30.72 \pm 0.05$~mag\\ 
$\textrm{RA}_{\textrm{SN}}$ &  &  & &  & & & &  &  $14^{\textrm{h}}32^{\textrm{m}}33.919^{\textrm{s}}$\\
$\textrm{Dec}_{\textrm{SN}}$ &  &  & &  & & & &  & $-44^{\circ}13'28.763^{\prime \prime}$\\
Stretch &  &  & &  & & & &  &  1.072 $\pm$ 0.014\\ 
$m_{B}^{\mathrm{peak}}$ &  &  & &  & & & &  & $11.11 \pm 0.05$~mag\\
$M_{B}^{\mathrm{peak}}$ &  &  & &  & & & &  & $-19.49 \pm 0.07$~mag\\

\hline
\end{tabular}
\end{center}
\centering
\label{table:Observations}
\end{table}

\begin{figure}

\begin{center}
  \includegraphics[width=0.45\textwidth]{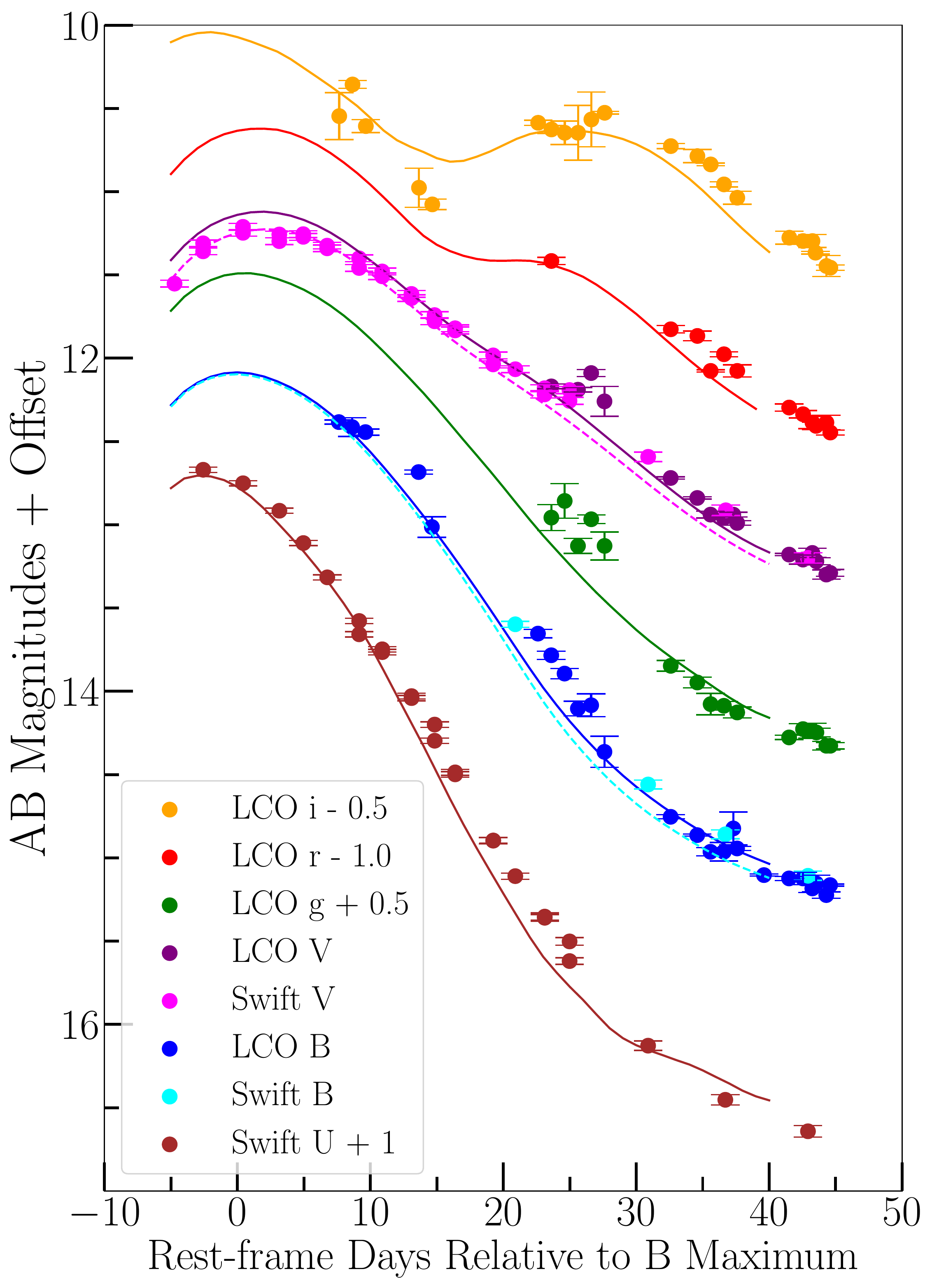}
  \caption{LCOGT and Swift light curves of SN\,2013aa around peak. LCOGT photometry in shown as B (blue), V (purple), g (green), r (red) and i (orange). Swift photometry shown as U (brown), B (cyan) and V (magenta). The photometry has been corrected for MW extinction. Solid and dashed lines are the \textsc{sifto} fits on the LCOGT and Swift photometry respectively.}
  \label{fig:sn2013aa_early_lightcurve}
\end{center} 

\end{figure}

SN\,2013aa was initially followed by the Las Cumbres Observatory
Global Telescope (LCOGT) Supernova Key Project \citep{Brown13}, with
the $B\!V\!gri$ light curves first published by \citet{Graham17arXiv}.
As mentioned in \citet{Graham17arXiv}, most of the near-peak
photometry was saturated, thus we complement the early-time light
curve with optical ($U\!B\!V$) data from the {\it Swift}
Optical/Ultraviolet Supernova Archive \citep[SOUSA;][]{Brown14}. This
data provides adequate coverage of the SN from $-10$ to $\sim$200~days
after peak.  Additionally, \citet{Graham17arXiv} present $gri$
photometry from the Gemini Multi-Object Spectrograph
\citep[GMOS;][]{Davies97}, at $\sim$400~days. In
Fig.~\ref{fig:sn2013aa_early_lightcurve}, we present the early-time
($-15$ to 50 days from peak) light curves of SN\,2013aa.

We fit the light curves with \textsc{sifto} \citep{Conley08}, with
which we recover a time of maximum light of $\mathrm{MJD_{max}} =
56342.69 \pm 0.18$, peak brightnss of $m_{B}^{\mathrm{peak}} = 11.11
\pm 0.05$~mag, peak color of $(B-V)_{0} = -0.03 \pm 0.05$~mag and a
stretch of $s = 1.072 \pm 0.014$. Restricting our fit to only
the {\it Swift} photometry, which covers the peak of the light curve, we calculate $s = 1.067 \pm
0.023$, consistent with what was found using all available data.

Adopting the distance modulus from SN\,2017cbv, $\mu = 30.72 \pm
0.05$~mag, SN\,2013aa had a $B$-band absolute magnitude at peak of
$M_{B}^{\mathrm{peak}} = -19.49 \pm 0.07$~mag.  The relatively high
peak absolute magnitude is consistent with its slightly broad light
curves.

\begin{figure}
\begin{center}
  \includegraphics[width=0.45\textwidth]{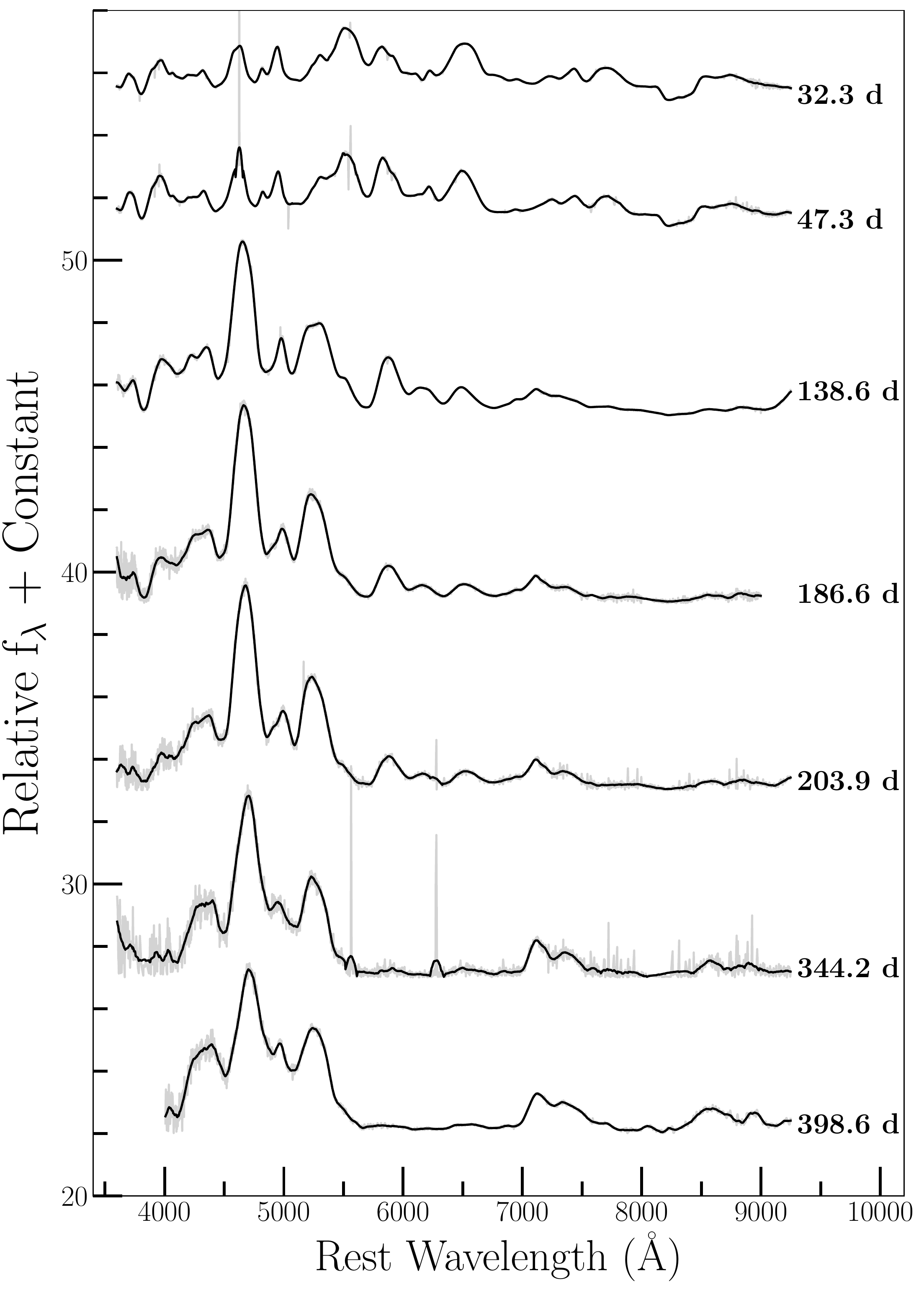}
  \caption{Spectra of SN\,2013aa. Raw spectra are shown in gray, smoothed spectra with black lines. }
  \label{fig:sn2013aa_spectra}
\end{center}
\vspace*{10mm}
\end{figure}

A collection of SN\,2013aa spectra is presented in
Fig.~\ref{fig:sn2013aa_spectra}, spanning from 32 to 398~days after
peak. These spectra have been published by \citet{Childress15} and
\citet{Graham17arXiv}. All the spectra were retrieved through the
WISeREP archive\footnote{\url{http://wiserep.weizmann.ac.il/}}
\citep{Yaron12}.

We used the Supernova Identification package
\citep[SNID;][]{Blondin07} and Superfit \citep{Howell05} at the
earliest spectrum (32.3 days after peak) to sub-classify the SN.  Both
packages reported SN~1991T-like objects as having the best-matching
spectra in accordance with the early-time light-curve evolution.
However, SN~1991T-like objects are difficult to distinguish from
lower-luminosity SNe~Ia a month after peak, and the sub-classification
is somewhat uncertain.  With this in mind, throughout this paper, we
will consider SN\,2013aa as a normal-to-overluminous SNe Ia. 
\\
\\

\subsection{HST Data}

\begin{figure*}[t!]

\begin{center}
	\includegraphics[width=0.9\textwidth]{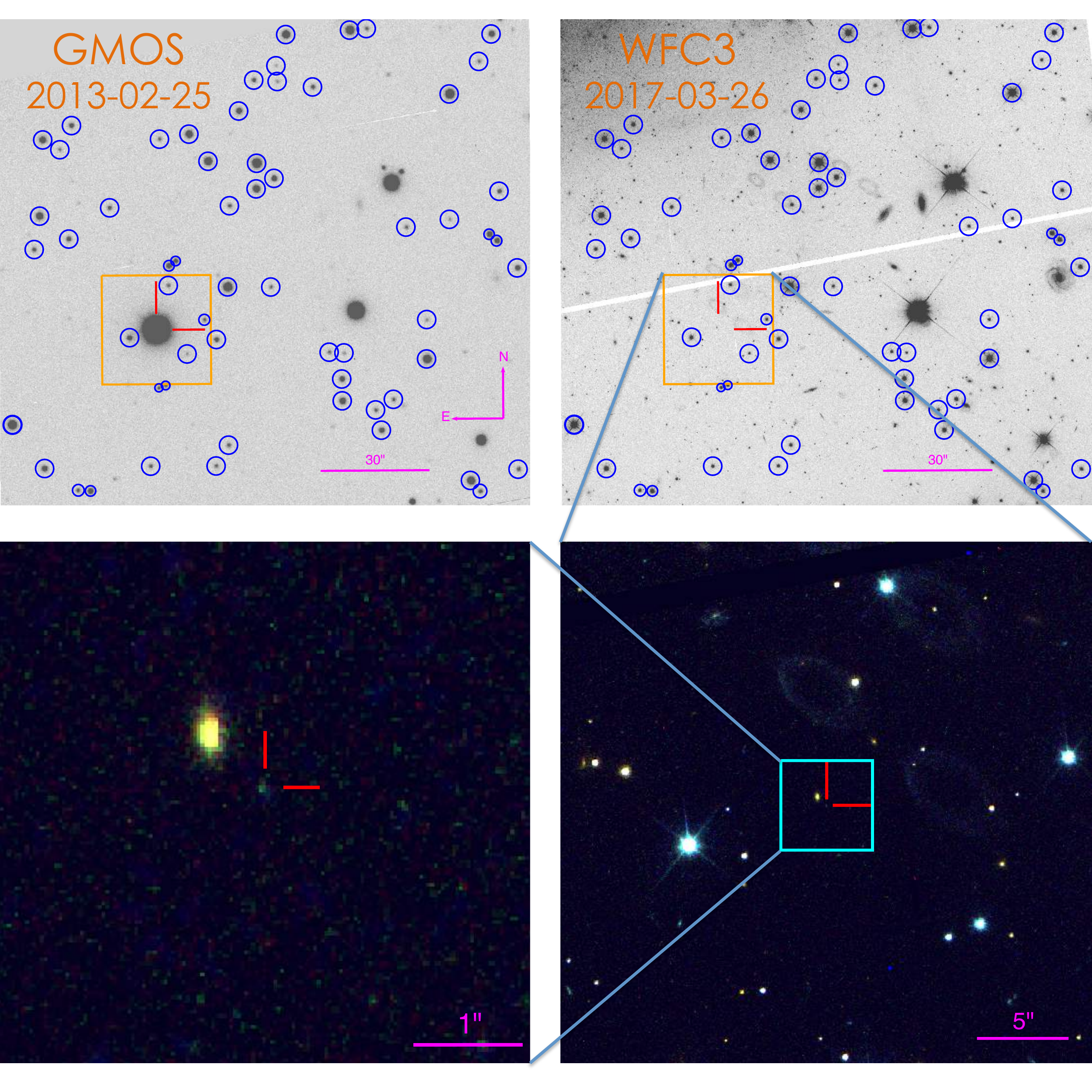}

	\caption{\textit{Top Images:} Explosion image taken by \textit{Gemini} with GMOS (left). \textit{HST} image of SN\,2013aa at 1500 day epoch taken with WFC3. Stars used for astrometric solution circled in blue. \textit{Bottom Images:} \textit{HST} RGB images centered on SN\,2013aa with $30^{\prime\prime}$ (right) and $5^{\prime \prime}$ (left) radii from the source. \label{fig:images_sn2013aa}}
\end{center}
\vspace*{5mm}
\end{figure*}

\label{subsec:very_late_data}
Due to its distance and significant offset from its host galaxy,
SN\,2013aa is an excellent target for late-time observations. Under
\textit{HST} program DD--14925 \citep{hstproposal}, we imaged SN\,2013aa
($\alpha = 14^{\textrm{h}}32^{\textrm{m}}33.919^{\textrm{s}}, \delta =
-44^{\circ}13'28.763^{\prime \prime}$) on 2017 March 22, 24, 26 \& 30
with the \textit{HST} Wide Field Camera 3 (WFC3). These observations
were obtained in parallel with STIS observations of SN\,2017cbv.  The
source was observed with wide-band filters \textit{F350LP},
\textit{F555W}, and \textit{F814W} at varying exposures times.
Photometric measurements are reported in
Table~\ref{tab:late_time_hst_photometry}.

We received \textit{HST} WFC3 image files in the FLC format, all of which have been corrected for
dark current, flat fielding, and charge transfer efficiency through
the \textit{HST} calibration pipeline.  We used the \textsc{IRAF}
package \textit{StarFind} to located reference stars for initial frame
alignment. We performed fine alignment of all images to one-another
using calibration algorithm \textsc{TweakReg}. With all frames
aligned, we ran the \textsc{AstroDrizzle} reduction package
\citep{astrodrizzle} for cosmic ray removal and generation of median
and drizzled science images for each \textit{HST} filter used. We
constructed a drizzled template image of all \textit{HST} filters by
overlaying each frame, which was then used as reference during
photometric calculations.

To determine the position of SN\,2013aa in the WFC images, we
determined a geometric transformation between the {\it HST} images and
Gemini images taken when the SN was brighter.  Using 21 stars common to each image and the Gaia stellar catalog, we calculated a WCS solution for both {\it HST} and Gemini images. We aligned the WCS of the \textit{HST} image to that of
\textit{Gemini} based on 72 common, unsaturated stars.  We then determine the position of SN\,2013aa in the \textit{HST} images.

We determined the positional systematic uncertainty related to our geometric transformation by performing the transformation many times using a bootstrap re-sampling (with replacement).
The final positional uncertainty is a combination of the
systematic uncertainty, the statistical uncertainty of the
geometric transformation, and the statistical uncertainty from centroiding the SN.

Our best estimate of the position of SN\,2013aa is $\alpha = 14^{\textrm{h}}32^{\textrm{m}}33.919^{\textrm{s}} \
\pm 0.003^{\textrm{s}} , \ \delta = -44^{\circ}13'28.76^{\prime\prime}
\ \pm 0.03^{\prime\prime}$. Images of SN\,2013aa with reference stars
are displayed in Figure ~\ref{fig:images_sn2013aa}. We detected a point source in our \textit{HST} image that was $+0.01^{\prime\prime}$ East and $+0.01^{\prime\prime}$ North of the supernova position found in the Gemini explosion image. This translates to a $0.31 \sigma$ offset in Right Ascension and a $0.33 \sigma$ offset in Declination. The position of the sources in both \textit{HST} and Gemini images agree with one another, which suggests that they are in fact the same source.

We performed Point Spread Function (PSF) photometry with
\textsc{DOLPHOT} \citep{dolphot} on the \textit{F350LP}, \textit{F555W}, and
\textit{F814W} images. \textsc{DOLPHOT} ran simultaneously on all
frames while using the combined template \textit{HST} frame for
reference. We used default WFC3 DOLPHOT parameters in the input file,
keeping the sigPSF value (minimum signal-to-noise for a PSF
calculation) at 10. Using 52 PSF stars in the photometric solution, we detected a point source in all three filter frames that was within the uncertainties of the astrometric solution, confirming that this was indeed SN\,2013aa. The source is shown most clearly in the bottom panel of Figure \ref{fig:images_sn2013aa}.

In this \textsc{DOLPHOT} detection, we measure the apparent magnitudes of SN\,2013aa to be $27.969 \pm 0.082$ in \textit{F350LP}, $27.971 \pm 0.280$ in \textit{F555W}, and $27.465 \pm 0.177$ in  \textit{F814W}, corresponding to signal-to-noise ratios of 13.3, 3.9, and 6.1, respectively. The brightness of this source is similar to that expected for a SN\,2013aa at this epoch.
We calibrated our apparent magnitudes
from \textsc{DOLPHOT} to AB magnitudes using the WFC3/UVIS2 photometry
zeropoint tables given by the Space Telescope Science Institute
(STSci)\footnote{\url{http://www.stsci.edu/hst/wfc3/analysis/uvis_zpts/uvis2_infinite/}}.
As a result of the default aperture correction performed by
\textsc{DOLPHOT} during the photometry calculation process, we applied
the infinite aperture zeropoint values to our generated absolute
magnitudes.

\begin{deluxetable*}{ccCccc}[t!]
\tablecaption{Photometric Observations \label{tab:late_time_hst_photometry}}
\tablecolumns{6}
\tablenum{2}
\tablewidth{0pt}
\tablehead{
\colhead{MJD} &
\colhead{Band} & \colhead{Exp. Time} & \colhead{AB Mag$^{a}$} & \colhead{Telescope} \\
& & (\mathrm{s}) & & &}
\startdata
57834 - 57842 & 350LP & 507 - 537 & 27.969 (0.082) & \textit{HST}/WFC3 \\
57834 - 57842 & 555W & 507-537 & 27.971 (0.280) & \textit{HST}/WFC3  \\
57834 - 57842 & 814W & 1014-1074 & 27.465 (0.177) & \textit{HST}/WFC3  \\
\enddata
\tablenotetext{a}{1-$\sigma$ uncertainties in parentheses.}
\tablecomments{Exposures were taken on 2017 March 22, 24, 26, and 30. All four days of observations were combined into a single image for each respective filter.}
\vspace*{-5mm}
\end{deluxetable*}

To determine the chance coincidence between SN\,2013aa and our
identified source, we look at other detected objects within a
$5^{\prime \prime}$ radius of 
SN\,2013aa.  We limit the sample of
reasonable objects to have ${\rm S/N} \ge 5$, be classified as a
star by Dolphot (type 1 or 2), have a roundness of $<$0.5 (as
determined by Dolphot), have a sharpness between $-0.3$ and 0.3, and
have a Dolphot photometric quality flag of 0 or 1.  We find 10
reasonable objects with a $5^{\prime \prime}$ radius, resulting in a
chance coincidence of only 0.2\%.

\section{Analysis} \label{sec:Analysis}

In this section we briefly detail how we generated a pseudo-bolometric
light curve from the photometric data described in
Section~\ref{sec:observations}. We then discuss our analysis of
different elemental decay chains responsible for light curve shape and
the process of determining each radioactive isotope mass based on the
fit to our bolometric luminosity data.

\subsection{Constructing a pseudo-bolometric light curve}
\label{subsec:constr_bol_lc}

In order to construct the pseudo-bolometric light curve of SN\,2013aa,
we employ similar techniques as performed for other late-time SN~Ia
studies, that includes the modification of the SN spectra to match a series of photometric observations and, subsequently, integration of these modified spectra over the optical wavelengths \citep[e.g.,][]{Graur16,dimitriadis17,graur17,Kerzendorf17, shappee17}. We correct all photometric data, both ground- and space based, for Milky Way Extinction according to \cite{cardelli89} with $R_v=3.1$, and find no host-galaxy extinction to correct for in the data.

For photometric epochs with phases of $\sim$100 to 200 days, we mangle \citep{Hsiao07} the closest-in-time spectrum to the LCOGT photometric data.  For the $\sim$400-day epoch, we perform the same operation with the GMOS photometry and spectrum.  For the 1500-day photometric epoch, there is no spectrum of SN\,2013aa or any other SN~Ia; instead, we use a 1000-day spectrum of SN\,2011fe \citep{Taubenberger15}. The bolometric flux is computed by integrating each modified synthetic spectrum from 4000 to 9000~\AA, obtaining errors by Monte Carlo re-sampling of the observed photometry. Finally, we calculate the optical bolometric luminosity by scaling the integrated flux with the distance to the SN, estimated in Section~\ref{subsec:early_data}.

The choice of wavelength range for generating the pseudo-bolometric
light curve was set by the wavelength coverage of the available
spectra, and in particular the GMOS spectrum (see
Fig.~\ref{fig:sn2013aa_spectra}).  While this wavelength range is
narrower than pseudo-bolometric light curves generated for other
SNe~Ia (usually $\sim$3500--10000 \AA), the dominant spectral lines of
SNe~Ia at these phases, mainly from iron peak elements and
\ion{Ca}{2}, are included in our wavelength range. We can estimate the
fraction of flux lost bluewards (3500--4000~\AA) and redwards
(9000--10000~\AA) of our pseudo-bolometric wavelength range by using
spectra of the well-observed SN\,2011fe: we calculate a fraction of
5\% and 9\% at 348d, reducing to 4\% and 7\% at 1034d. Our closest spectrum to the GMOS spectrum at 398d is the WiFeS spectrum at 344d, which spans from 3500--9280\AA, for which we estimate a fraction of flux lost bluewards (i.e. the integrated flux from 3500--4000 over the integrated flux from 3500--9280\AA) and redwards (i.e. the integrated flux from 9000--9280 over the integrated flux from 3500--9280\AA) of our pseudo-bolometric wavelength range to be 1.2\% and 1.5\% respectively. By using spectra of the well-observed SN\,2011fe, which cover a wider wavelength range (3000--10000\AA), the equivalent blueward-redward flux losses are 5\% and 9\% at 348, reducing to 4\% and 7\% at 1034d.
\vspace*{-1mm}

\subsection{The bolometric light curve model}
\label{subsec:bol_lc_model}

\begin{figure*}[t!]

\centering
\gridline{\fig{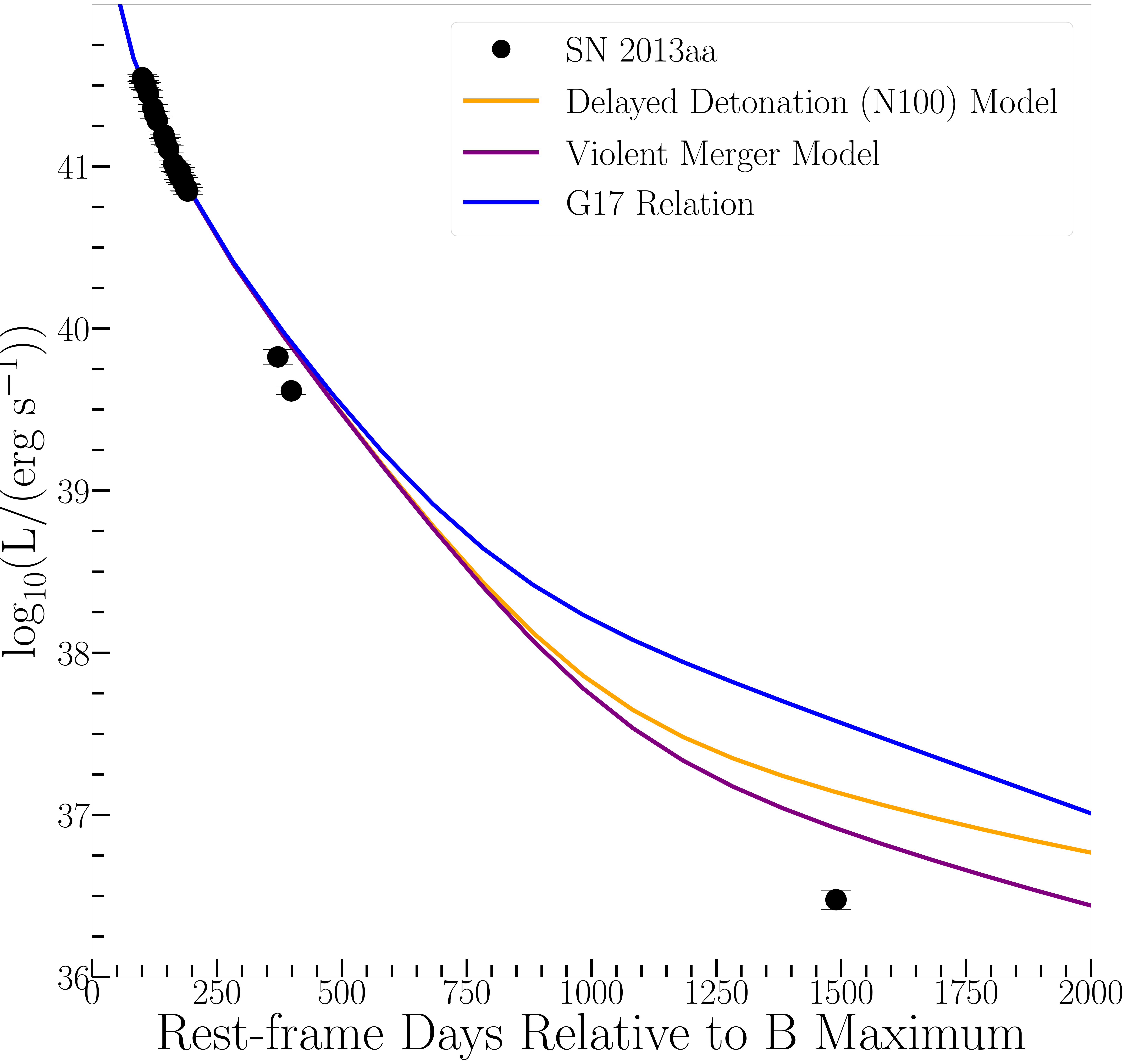}{0.5\textwidth}{}
\fig{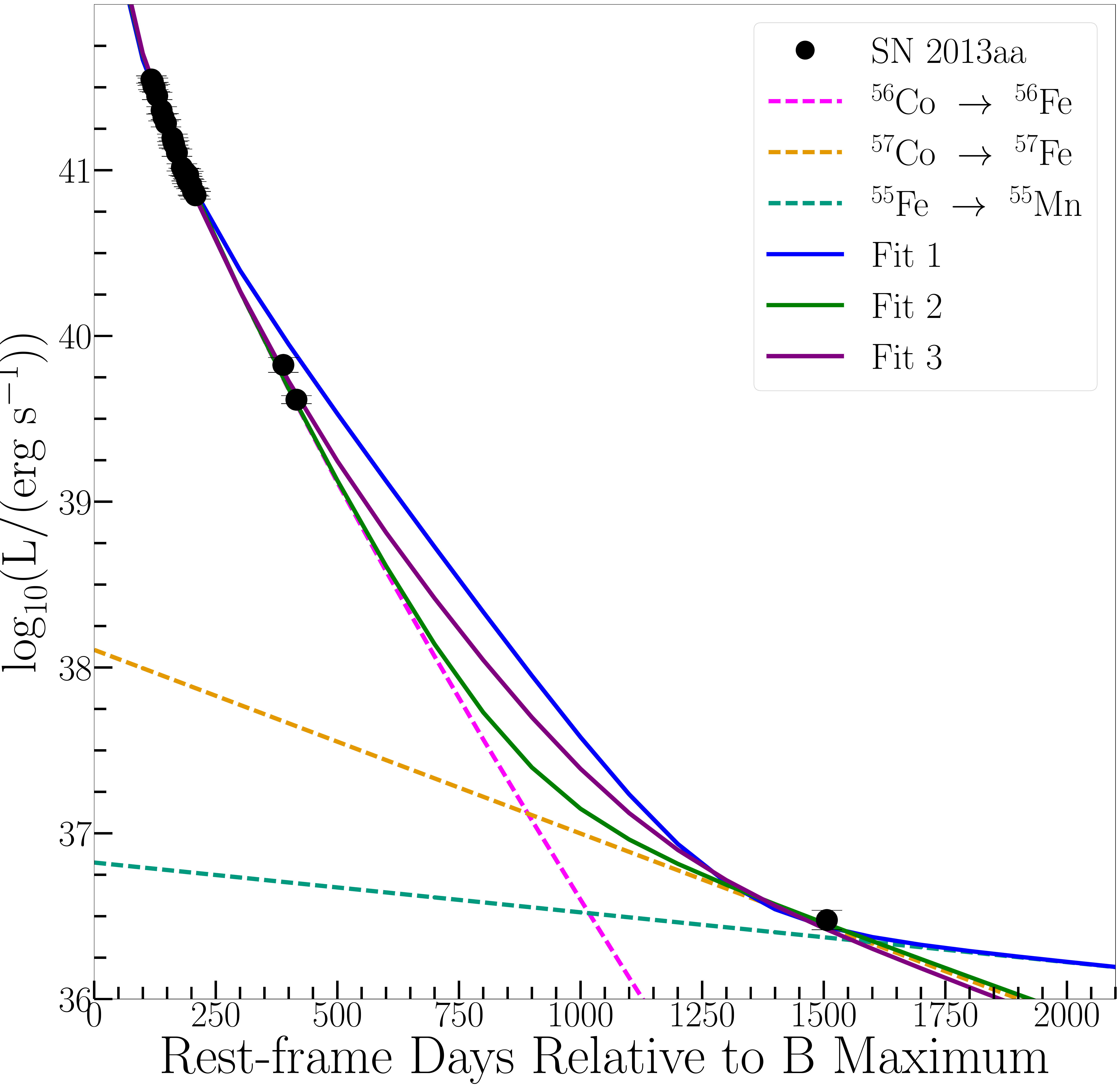}{0.49\textwidth}{}
}
\vspace*{-5mm}
\caption{\textit{Left:} Bolometric luminosities of SN\,2013aa with respect to SD (orange) and DD (purple) explosion models using fractions of ${}^{56}\textrm{Co}$ predicted by \citet{ropke12}. Blue line is calculated from the expected mass of ${}^{57}\textrm{Co}$ based on fit for ${}^{57}\textrm{Co}$ / ${}^{56}\textrm{Co}$ versus stretch shown by \citet{graur17}. This trend is plotted in Figure \ref{fig:stretch}. \textit{Right:} Our three separate fits to bolometric data. Values found for ${}^{56}\textrm{Co}$, ${}^{57}\textrm{Co}$, and ${}^{55}\textrm{Fe}$ reported in Table \ref{tab:models}.  Other three lines represent the decomposition of ${}^{56}\textrm{Co}$, ${}^{57}\textrm{Co}$, and ${}^{55}\textrm{Fe}$ decay chains using the masses found in Fit 2 as well as the upper limit of $M({}^{55}\textrm{Fe})$ measured in Fit 1. \label{fig:bol_plot} }
\vspace*{8mm}
\end{figure*}

The light curve of a SN Ia is powered by the thermalization of the expanding ejecta due to the deposition of energy from the radioactive decay of several decay chains. At early times, the dominant contribution comes from ${}^{56}\textrm{Ni}$, the most abundant synthesized element, and its daughter isotope, ${}^{56}\textrm{Co}$, with its decay channel $\ce{{}^56Co ->[ t_{1/2}=77.2 \textrm{d} ] {}^56Fe }$ being the most important for epochs up to 2~yrs after explosion. At later times, and as the column density of the expanding ejecta decreases, additional energy is deposited by the radioactive decays of $\ce{{}^57Co ->[ t_{1/2}=271.2 \textrm{d} ] {}^57Fe }$ and \ce{{}^55Fe ->[ t_{1/2}=999.67 \textrm{d} ] {}^55Mn }. All of these decay chains produce $\gamma$-rays, X-rays and charged leptons (positrons, Auger electrons, and internal conversion electrons). In our analysis we employ the same decay energies and constants as presented in Table 2 of \cite{seitenzahl14}. In this framework, the luminosity produced can be approximated by the Bateman equation:
\vspace*{1mm}
\begin{equation}\label{eq:eq4}
\begin{split}
&L_A(t) = 2.221 \frac{\lambda_A}{A} \frac{M(A)}{M_{\odot}} \frac{q^x_A + q^l_Af^{l}_A(t) +  q^{\gamma}_Af^{\gamma}_A(t)}{\mathrm{keV}}\\
& \ \ \ \ \ \ \ \ \ \ \ \ \ \ \ \ \ \ \ \ \ \ \ \ \ \ \ \ \ \ \textrm{exp}(-\lambda_{A}t) \ \times \ \ 10^{43} \ \textrm{erg} \ \textrm{s}^{-1}
\end{split}
\end{equation}

where $t$ is time since explosion, $\lambda_A$ is the decay constant, $A$ is the atomic number, and $q^l$, $q^\gamma$, and $q^x$ are the average energies of charged leptons, $\gamma$-rays, and X-rays, respectively, per decay. In this equation, $f^{\gamma}_A(t)$ and $f^{l}_A(t)$ describe the trapping of the deposited energy of the $\gamma$-rays and charged leptons respectively, and, assuming homologous expansion, are given by

\begin{equation}\label{eq:eq5}
f^{\gamma,l}_{A} = 1- \textrm{exp}\Big[-\Big(\frac{t^{\gamma,l}_{A}}{t}\Big)^2 \Big]
\end{equation}

In previous late-time studies, such as \citet{Graur16},
\citet{shappee17} and \citet{graur17}, with late-time data $>500$
days, the authors consider only the charged leptons deposited energy, for which they assume complete trapping in the decay of ${}^{56}\textrm{Co}$ (i.e., $f^{l}_{A} = 1$), and no positron trapping in the decays of ${}^{57}\textrm{Co}$ and ${}^{55}\textrm{Fe}$ (i.e., $f^{l}_{A} = 0$). For the
${}^{56}\textrm{Co}$ $\gamma$-rays, a timescale of $t^{\gamma}_{56}
\approx 35$ days was found to fit the late-time light curves of
several SNe~Ia \citep{sollerman04,Stritzinger07,leloundas09,zhang16} While these SNe~Ia do have lower predicted mass of ${}^{56}\textrm{Ni}$ than SN~2013aa, the application of $t^{\gamma}_{56}
\approx 35$ days is still an adequate assumption and has no effect on the analysis.

While Equation~\ref{eq:eq4} describes the \textit{bolometric}
luminosity (that is, the complete energetic output across the
electromagnetic spectrum), the photometric data presented here and in
(most of) the aforementioned studies are primarily optical data, with
some cases including near-infrared observations. A common approach is
to assume that the \textit{optical} luminosity scales with the
complete bolometric one as $L_{\rm opt}(t) = B(t) \times L_{\rm
  bol}(t)$, where $B(t)$ is the fraction of the bolometric luminosity
in the optical and is often assumed to be a constant in time. In this sense, 1/$B(t)$ resembles a ``bolometric correction'', i.e. a function that transforms the optical flux to a bolometric one. We can
estimate $B(t)$ by calculating the ratio between the
${}^{56}\textrm{Co}$ mass found by fitting the late-time data with Equation \ref{eq:eq4} over the total ${}^{56}\textrm{Ni}$ mass as determined from data around peak (where ${}^{56}\textrm{Ni}$ dominates), for which the non-optical contribution at this phase is $\leq$15\% \citep[e.g. see][for SN\,2011fe]{Pereira13}. Values of $B(t)$ calculated by \citet{graur17} for a sample of SNe Ia with late-time data range from  20-40\%.
However, \citet{dimitriadis17} showed that, for
SN\,2011fe, a non-constant $B$ can explain the increase of the
late-time non-optical contribution, approximating the optical
contribution with a sigmoid function:

\begin{equation}\label{eq:eq6}
  B(t) = 1-\frac{P_{0}}{1-e^{P_{1} \times (t-P_{2})}}
\end{equation}

In that work, this non-optical contribution, consisting of the
$J\!H\!K$ near-infrared bands, increases from $\sim$5 to 35\%, from
200 to 600 days after the $B$-band maximum brightness. This effect can
be seen as a faster decline of the (optical bolometric) light curve at
these epochs, compared to the expected radioactive decay slope,
predicted by known radioactive decay chains. The physical origin
of this faster decline remains elusive: positron escape models, a
re-distribution of optical flux to the mid/far-infrared
\citet{Fransson15} or time-dependent effects, such as freeze-out could
provide an explanation.

\subsection{Results from Light-Curve Model Fitting}
\label{subsec:results}

\begin{deluxetable*}{cccccccc}
\centering
\tablecaption{Model Fit to Pseudo-Bolometric Light Curve Data\label{tab:models}}
\tablecolumns{8}
\tablenum{4}
\tablewidth{0.99\textwidth}

\tablehead{
\colhead{Model} &
\colhead{${}^{56}\textrm{Co}$} & \colhead{${}^{57}\textrm{Co}$} & \colhead{${}^{55}\textrm{Fe}$} & \colhead{$t^{\gamma}_{56}$}& 
\colhead{$t^{l}_{56}$}& 
\colhead{$\chi^2$} & \colhead{DOF} \\
& $(\textrm{M}_{\odot})$ & $(\textrm{M}_{\odot})$ & $(\textrm{M}_{\odot})$ &  (days from explosion) & (days from explosion)}
\startdata
Fit 1 &  $0.589^{+0.014}_{-0.014}$ & $0.00002^{+0.0001}_{-0.00002}$ & $0.006^{+0.001}_{-0.006}$ & $35^{a}$ & - & 298.4 & 23 \\
Fit 2 & $0.631^{+0.015}_{-0.015}$ & $0.006^{+0.001}_{-0.006}$ & $0.0002^{+0.0007}_{-0.0002}$ & $35^{a}$ & $281.02^{+16.44}_{-15.29}$ & 20.4 & 22 \\
Fit 3 & $0.732^{a}$ & $0.015^{+0.0075}_{-0.015}$ & $0.0000007^{+0.000007}_{-0.0000007}$ & $35^{a}$ & - & 21.3 & 20 \\
Fit 4 & $0.59^{+0.01}_{-0.01}$ & $0.006^{+0.001}_{-0.001}$ & $0^{a}$ & $35^{a}$ & - & 299.7 & 24 \\
\enddata

\tablenotetext{a}{Fixed during fitting.}

\end{deluxetable*}

In this work, we will explore four models for the late-time light
curve of SN\,2013aa: (1) Complete positron trapping (i.e., $f^{l}_{A} =
1$ \& negligible $t^{l}_{55, 56, 57}$) and free-streaming $\gamma$-rays at late-times (i.e., $f^{\gamma}_{55,57} =
0$ \& negligible $t^{\gamma}_{55, 57}$) , (2) the same as 1, but with possible positron escape, for which we will assume a same form
of $f^{l}_{A}$ as the trapping function of the $\gamma$-rays (i.e., as
in Equation~\ref{eq:eq5}), (3) the same as 1, but with a time-dependent non-optical
contribution (Equation~\ref{eq:eq6}), and (4) the same as 1, but with no ${}^{55}\textrm{Fe}$, as was assumed by \citet{graur17}. For all of our fits, we 
assume $t^{\gamma}_{56} = 35$~days, and, by applying a Markov-Chain
Monte-Carlo fitting algorithm, determine the amount of
${}^{56}\textrm{Co}$, ${}^{57}\textrm{Co}$, and ${}^{55}\textrm{Fe}$.
In our analysis, we use \textsf{emcee}, a Python-based application of
an affine invariant MCMC with an ensemble sampler
\citep{foreman-mackey13}. Working with an MCMC allows for the
detection of degeneracy amongst free variables that could not be
properly identified with a standard $\chi^{2}$ fitting algorithm. Unfortunately, SN\,2013aa has no data between 400 and 1500~days.  As a
result, it is difficult to separate the contributions of
${}^{57}\textrm{Co}$ and ${}^{55}\textrm{Fe}$ to the late-time light
curve.  However, the current data are still constraining for explosion
models. 

An important step in consistently comparing our late-time mass
estimates of the different scenarios considered, is an accurate
determination of the total ${}^{56}\textrm{Ni}$ mass, synthesized in
the explosion. At early times, the luminosity is dominated by the
${}^{56}\textrm{Ni}$ decay and almost all of the light is emitted in
the optical \citep[e.g.,][]{Pereira13}. In the sample study of
\citet{graur17}, the authors estimate the ${}^{56}\textrm{Ni}$ mass by
fitting a straight line to the $M_{56}$ values of \citet{Childress15}
over their \textsc{sifto} stretch values. A similar calculation for
SN\,2013aa yields $M_{56} = 0.732 \pm 0.151$~M$_{\odot}$. As a
consistency check, we additionally estimate the ${}^{56}\textrm{Ni}$
mass from the bolometric luminosity at peak, following the widely-used
Arnett law \citet{arnett82}.  Using our early-time photometry
(Section~\ref{subsec:early_data}) and a template SN~Ia spectrum from
\citet{Hsiao07} at peak, we integrate the spectrum and estimate
$L_{peak} = 1.56 \pm \: 0.05 \times 10^{43} \: \mathrm{erg\:s^{-1}}$.
Assuming a rise time of 17~days, we estimate $M_{56} = 0.73 \pm 0.03
\: \mathrm{M_{\odot}}$. In the following sections, we will follow the
\citet{graur17} approach and adopt $M_{56} = 0.732 \pm 0.151 \:
\mathrm{M_{\odot}}$.

In Fit 1, we considered complete positron
trapping and a fixed $t^{\gamma}_{56} = 35$ days in fitting for the masses of ${}^{56}\textrm{Co}$, ${}^{57}\textrm{Co}$, and ${}^{55}\textrm{Fe}$.
We find a ${}^{56}\textrm{Co}$ mass of $0.589^{+0.0140}_{-0.0140} \ \textrm{M}_{\odot}$, which is 20\% less than the total mass of ${}^{56}\textrm{Ni}$ calculated from the near-peak data. Additionally, we find estimates of ${}^{57}\textrm{Co}$ and ${}^{55}\textrm{Fe}$ masses of $M({}^{57}\textrm{Co}) = {2 \times 10^{-5}}^{+{1 \times 10^{-4}}}_{-{2 \times 10^{-5}}}$~M$_{\odot}$, and $M({}^{55}\textrm{Fe})=0.006^{+0.001}_{-0.006} \ \textrm{M}_{\odot}$. This fit yields a $\chi^2/{\rm dof} = 298.4/23$, and we calculate a mass ratio of ${}^{57}\textrm{Co}/{}^{56}\textrm{Co} = {3 \times 10^{-5}}^{+{2 \times 10^{-4}}}_{-{3 \times 10^{-5}}}$. For this scenario, the best-fitting values
have significantly more ${}^{55}\textrm{Fe}$ than
${}^{57}\textrm{Co}$, although the range of allowed values include
having the mass hierarchy inverted.

Unlike the other fits displayed in Figure \ref{fig:bol_plot}, Fit 1 is significantly more luminous at 400~days than the data, suggesting that -- under the assumption of a constant bolometric correction -- incomplete positron trapping occurs at 400~days, and is therefore likely to also occur at later times.
  
In Fit 2, we fit for all three radioactive isotope masses in addition
to $t^{l}_{56}$, which allows for positron leakage (see
Equation~\ref{eq:eq5}). This varies from Fit 1 in that we now consider only partial positron trapping as well as a fixed $t^{\gamma}_{56} = 35$ days. The ${}^{56}\textrm{Co}$, ${}^{57}\textrm{Co}$, and ${}^{55}\textrm{Fe}$ masses are estimated to be $0.631^{+0.0150}_{-0.0150} \ \textrm{M}_{\odot}$,  $0.006^{+0.001}_{-0.006} \ \textrm{M}_{\odot}$, and $0.0002^{+0.0007}_{-0.0002} \ \textrm{M}_{\odot}$, respectively.

We find that the best-fitting value of ${}^{56}\textrm{Co}$ is 14\% less than the near-peak estimate of ${}^{56}\textrm{Ni}$, and we
calculate a mass ratio ${}^{57}\textrm{Co}/{}^{56}\textrm{Co} =
0.01^{+0.002}_{-0.01}$. This model has a $\chi^{2}/{\rm dof} = 20.4/22$.  Fit 2 is much better at matching the data near 400~days than Fit 1.  We find that fitting for partial rather than complete positron trapping yields a timescale of $t^{l}_{56}=281.02^{+16.440}_{-15.290}$ days for lepton escape. 

In Fit 3, we fit for ${}^{57}\textrm{Co}$,
${}^{55}\textrm{Fe}$, and each free parameter of the sigmoid function
in Equation~\ref{eq:eq6}, while fixing $M_{56}$ to the value
determined from the early-time data, $M_{56} = 0.732 \pm 0.151 \:
\mathrm{M_{\odot}}$. Similar to Fit 1, this model includes complete positron trapping, but with an increasing non-optical contribution to the total luminostiy of the light curve. We measure the ${}^{57}\textrm{Co}$ mass to be $0.015^{+0.0075}_{-0.015} \: \textrm{M}_{\odot}$, and a mass ratio ${}^{57}\textrm{Co}/{}^{56}\textrm{Co} =
0.02^{+0.01}_{-0.02}$. The best-fitting value for the mass of
${}^{55}\textrm{Fe}$ is only ${7 \times 10^{-7}}^{+{7 \times 10^{-6}}}_{-{7 \times 10^{-7}}}$~M$_{\odot}$,
significantly smaller than the best-fitting values of the other fits,
but consistent with their range for the ${}^{55}\textrm{Fe}$ mass. This fit has a $\chi^{2}/{\rm dof} = 21.3/20$ and, like Fit 2, matches the data at 400 days \ref{fig:bol_plot}.

Finally for Fit 4, we set the ${}^{55}\textrm{Fe}$ mass to be zero.  This is done to be consistent with the \citet{graur17} analysis.  We find ${}^{56}\textrm{Co}$ and ${}^{57}\textrm{Co}$ masses of $0.59^{+0.010}_{-0.010} \ \mathrm{M_{\odot}}$ and $0.006^{+0.001}_{-0.001} \ \mathrm{M_{\odot}}$, respectively. The best-fitting value of ${}^{56}\textrm{Co}$ is $20\%$ less than the total the near-peak estimate of ${}^{56}\textrm{Ni}$, and we calculate a mass ratio ${}^{57}\textrm{Co}/{}^{56}\textrm{Co} =
0.01^{+0.002}_{-0.002}$. This fit has a $\chi^{2}/{\rm dof} = 299.7/24$ and, like Fit 1, is over-luminous, relative to the data, around 400~days after peak brightness. Although Fit 4 is not a particularly good representation of the data, we use the mass ratios measured here when comparing to other SNe~Ia examined by \citet{graur17} in Section~\ref{subsec:comparison}.

Best-fitting parameters for each model are reported in Table~\ref{tab:models}, with each respective fit plotted in Figure~\ref{fig:bol_plot}.

\begin{deluxetable*}{lccccccc}
\centering
\tablecaption{Explosion Model Characteristics\label{tab:explosion_models}}
\tablecolumns{8}
\tablenum{5}
\tablewidth{0.99\textwidth}

\tablehead{
\colhead{Model} &
\colhead{Description} & \colhead{Density} & \colhead{M(WD)} & \colhead{M($^{56}$Ni)}& 
\colhead{$^{57}\textrm{Co}/^{56}\textrm{Co}$}& 
\colhead{Stretch} & \colhead{Ref.}\\
& & (g $\textrm{cm}^{-3}$) & $(\textrm{M}_{\odot})$ &  $(\textrm{M}_{\odot})$ & &}
\startdata
\textbf{Single Degenerate} &  & & & &  &  &  \\ 
W7 &  Deflagration & $2 \times 10^{9}$ & $1.38$ & $0.59$ & 0.041 & 0.90 & \citealt{iwamoto99} \\
ddt$\_$n100 & Delayed Detonation & $2.9 \times 10^{9}$ & 1.40 & 0.60 & 0.031 & 0.83 & \citealt{seitenzahl13} \\
det$\_$1.06 & Detonation & $4.2 \times 10^{7}$ & $1.06$ & $0.56$ & 0.006 & 0.75 & \citealt{Sim10ApJ} \\
doubledt$\_$CSDD-S & Double Detonation & $8.5 \times 10^{6}$ & $0.79$ & $0.21$ & 0.044 & 0.89 & \citealt{Sim12MNRAS} \\
def$\_$N100def & Pure Deflagration & $2.9 \times 10^{9}$ & $1.40$ & $0.36$ & 0.038 & 0.84 & \citealt{Fink14MNRAS} \\
det$\_$ONe15e7 & O-Ne WD Detonation & $1.5 \times 10^{8}$ & $1.23$ & $0.96$ & 0.009 & 0.79 & \citealt{Marquardt15AA} \\
gcd$\_$GCD200 & Detonation & $1.0 \times 10^{6}$ & $1.40$ & $0.74$ & 0.025 & 0.92 & \citealt{Seitenzahl16AA}\\
\hline 
\textbf{Double Degenerate} &  & & & &  &  &  \\
merger$\_$11+09 & Violent Merger & $2.0 \times 10^{6}$ & $1.10+0.90$ & $0.10$ & 0.024 & 1.03 & \citealt{pakmor12} \\
merger$\_$09+09 & Violent Merger & $3.8 \times 10^{6}$ & $0.90+0.90$ & $0.10$ & 0.003 & 0.80 & \citealt{pakmor10} \\
merger$\_$09+076$\_$Z1 & Violent Merger & $2.0 \times 10^{6}$ & $0.90+0.76$ & $0.18$ & 0.009 & 0.89 & \citealt{Kromer13ApJ} \\
merger$\_$09+076$\_$Z0.01 & Violent Merger & $2.0 \times 10^{6}$ & $0.90+0.76$ & $0.18$ & 0.003 & 0.99 & \citealt{Kromer16MNRAS} \\
\enddata

\end{deluxetable*}

\section{Discussion} \label{sec:Discussion}

\subsection{Comparison to Explosion Models}

The mass ratios between given radioactive isotopes are indicators of the explosion mechanism in SNe~Ia. We compare our values for ${}^{57}\textrm{Co}$/${}^{56}\textrm{Co}$ with the two explosion models presented in \citet{ropke12}, both of which probe the two extremes of the density at the location of the explosion: The ddt$\_$n100 \citep{seitenzahl13}, a Delayed Detonation, near-Chandrasekhar mass explosion model, with $\rho\sim${}$3 \times 10^{9} \ \textrm{g cm}^{-3}$, and the merger$\_$11+09 \citep{pakmor12}, a Violent Merger model of 1.1 and 0.9 $\textrm{M}_{\odot}$ WDs, with $\rho\sim${}$2 \times 10^{6} \ \textrm{g cm}^{-3}$. Figure~\ref{fig:bol_plot} illustrates that the Violent Merger model has a 1500-day luminosity that is more similar to that of SN\,2013aa than the Delayed Detonation model. However, both models predict a significantly more luminous event than SN\,2013aa.  Our preferred description of the data (Fit 3) has ${}^{57}\textrm{Co}$/${}^{56}\textrm{Co} = 0.02^{+0.01}_{-0.02}$, which is more than 0.4$\sigma$ below that of the Violent Merger model (${}^{57}\textrm{Co}$/${}^{56}\textrm{Co} = 0.0242$) and 1.1$\sigma$ below that of the Delayed Detonation model (${}^{57}\textrm{Co}$/${}^{56}\textrm{Co} = 0.0311$). The other scenarios described in Section~\ref{subsec:results} have even smaller ratios of ${}^{57}\textrm{Co}$/${}^{56}\textrm{Co}$.

Despite the best-fitting values being consistent with zero, the parameter space of our model fits provides estimates for the abundances of ${}^{57}\textrm{Co}$ and ${}^{55}\textrm{Fe}$ at this late-time epoch. Due to the difficulty in detecting ${}^{55}\textrm{Fe}$, other late-time studies have constrained this isotopic abundance based on the ratio of ${}^{57}\textrm{Co}/{}^{55}\textrm{Fe}$ predicted in SD and DD explosion models such as \citet{ropke12}, \citet{ohlmann14} and \citet{iwamoto99}. We, however, find it inconsistent to
enforce a ratio of ${}^{57}\textrm{Co}/{}^{55}\textrm{Fe}$, but not
that of ${}^{57}\textrm{Co}/{}^{56}\textrm{Co}$ in fitting for the
abundance of all radioactive isotopes. In our three fits we choose not to constrain the
mass of ${}^{57}\textrm{Co}$ nor that of ${}^{55}\textrm{Fe}$, and thus explore the parameter space of each fit without the confinement of an explosion model mass ratio. Nonetheless, the degeneracy between the masses of ${}^{57}\textrm{Co}$ and ${}^{55}\textrm{Fe}$ cannot be broken by our limited late-time data, and ultimately requires future
observation of SN\,2013aa in epochs where the presence of
${}^{55}\textrm{Fe}$ becomes more
prominent in the bolometric light curve.

\subsection{Comparison to Other Supernova Observed at Late-Time Epochs}  \label{subsec:comparison}

\begin{figure*}[t!]
\vspace*{-5.5mm}
\centering
\gridline{\fig{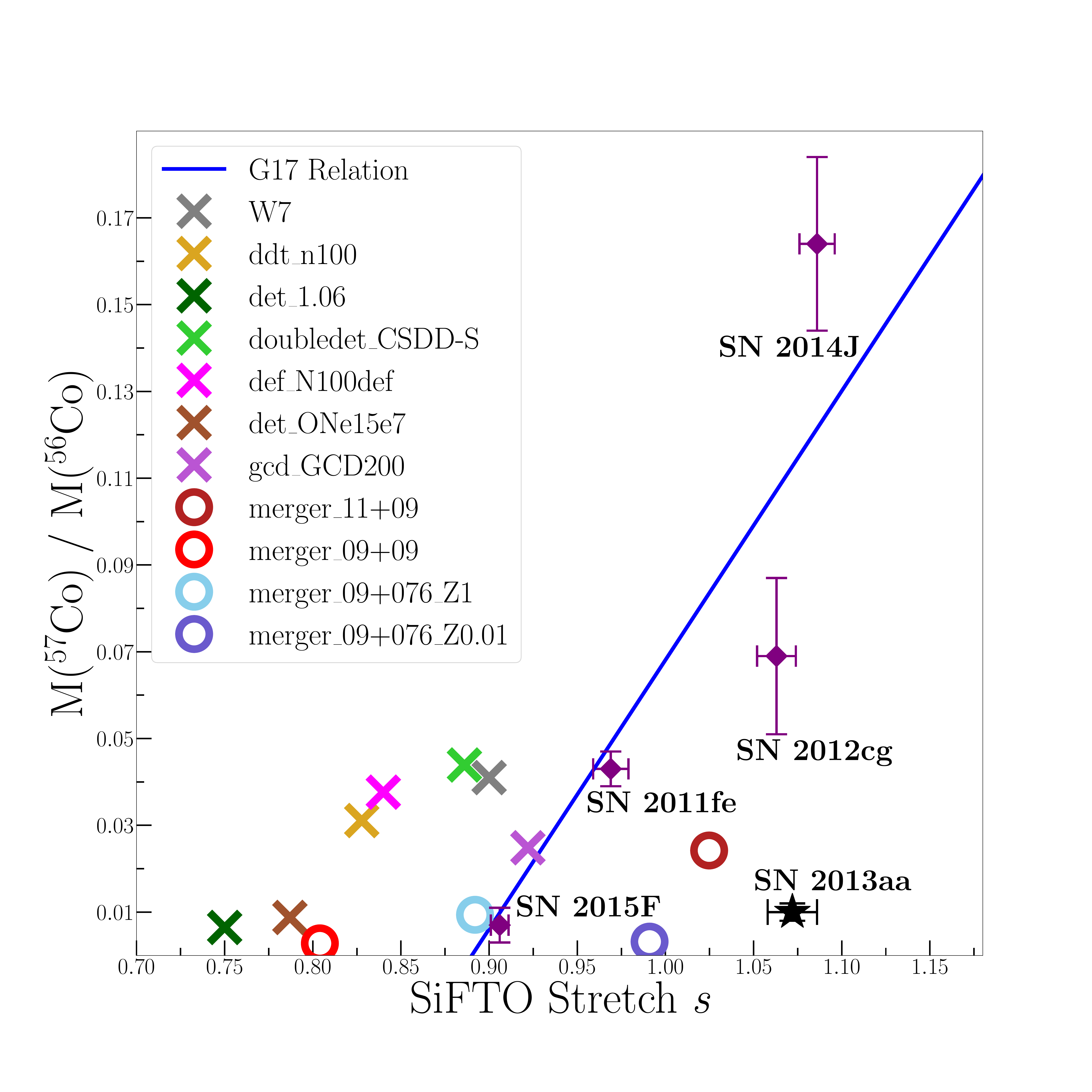}{0.5\textwidth}{}
          \fig{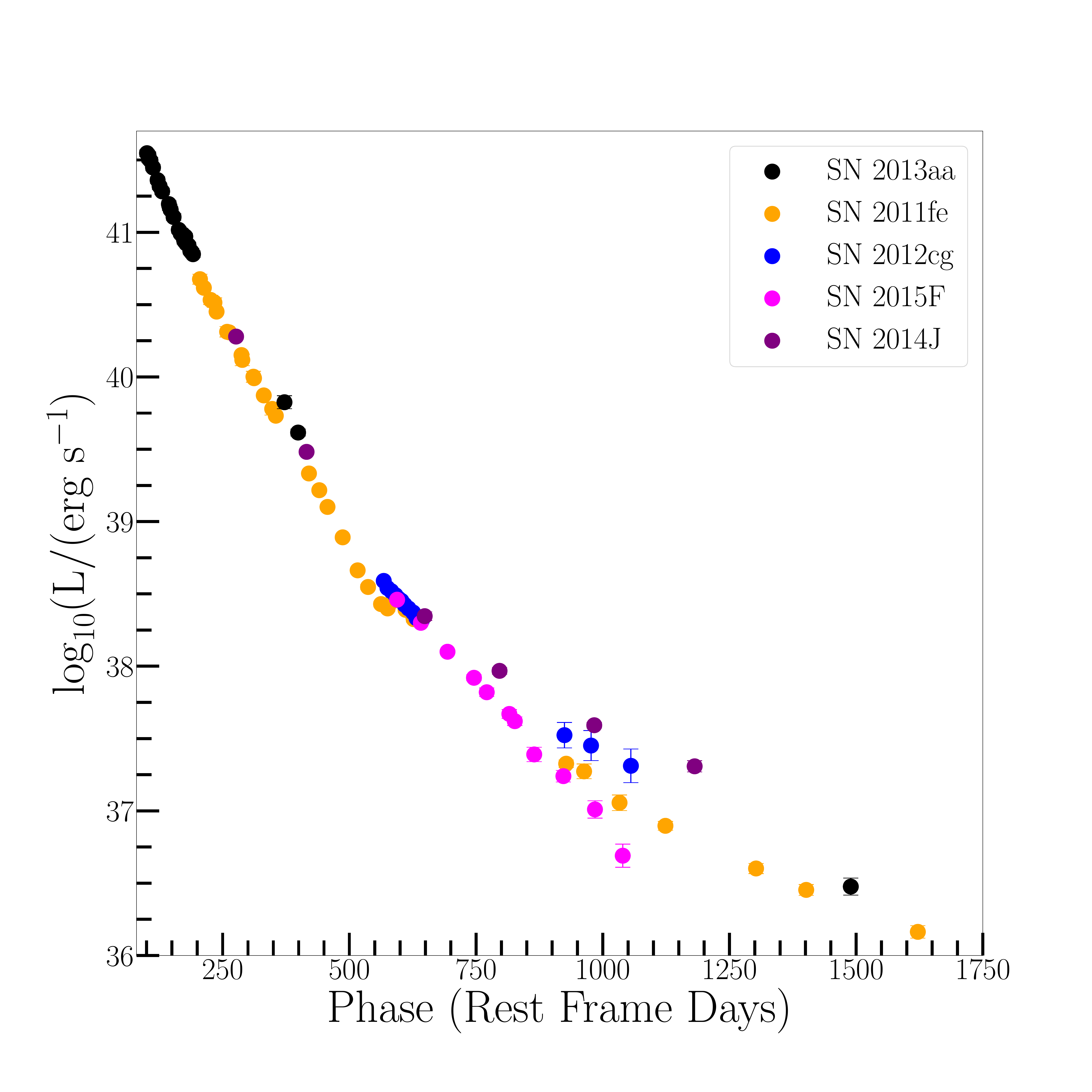}{0.5\textwidth}{}}
\vspace*{-10mm}
\caption{{\textit{Left:} Reproduction of mass ratio vs. stretch plot from \citet{graur17} with added SN\,2013aa point. We displayed the ratios of a variety of single-degenerate and double-degenerate explosion models with their predicted stretch value provided by the Heidelberg Supernova Model Archive. Blue line represents our fit to the \citet{graur17} data, excluding SN\,2013aa in the fit. \textit{Right:} Combined pseudo-bolometric light curve data from all late-time SNe Ia studies. } 
\label{fig:stretch}}
\vspace*{10mm}
\end{figure*}

SN\,2013aa is the fifth SN~Ia used to constrain
explosion models via mass ratios of late-time decay elements. Using the four
SNe~Ia with previous extremely late-time photometry, \citet{graur17}
found a linear trend between light-curve shape (specifically,
\textsc{sifto}-calculated stretch values) and
M(${}^{57}$Co)/M(${}^{56}$Co).  They also found a linear trend between
the change in pseudo-bolometric luminosity between 600 and 900 days
($\Delta\textrm{L}_{900} =
\log_{10}(\textrm{L}_{600}/\textrm{L}_{900})$
and the time at which freeze-out effects are most
prevalent in the light curve, $\textrm{t}_{\textrm{freeze}}$. We
reproduce these trends in Figure~\ref{fig:stretch} by fitting a line
to the four original data points. In Figure~\ref{fig:stretch}, we also include the values for SN\,2013aa found in Fit 4, in which we fit the light curve only for $M({}^{56}\textrm{Co})$ and $M({}^{57}\textrm{Co})$, with $M({}^{55}\textrm{Fe})=0$. This fit has the same assumptions as the \citealt{graur17} analysis (i.e. complete positron trapping). From the figure, we
see that SN\,2013aa is a large outlier to the \citet{graur17} trend.

Using the \citet{graur17} relation, we estimate a theoretical mass
ratio of M(${}^{57}$Co)/M(${}^{56}$Co) corresponding to the stretch
value we find for SN\,2013aa. We plot that model with respect to
bolometric luminosity data as the blue line in Figure
\ref{fig:bol_plot}.  The \citet{graur17} relation predicts a
luminosity that is more than an order of magnitude above that of
SN\,2013aa.  We conclude that either SN\,2013aa is extremely abnormal
or the \citet{graur17} relation does not hold for a larger sample.

We also use Figure~\ref{fig:stretch} to explore the implication of various explosion models abundances. Apart from the already discussed ddt$\_$n100, merger$\_$11+09 and W7, we include models from the Heidelberg Supernova Model Archive (HESMA)\footnote{\url{https://hesma.h-its.org/doku.php?id=start}} that include various binary configurations and explosion mechanisms. We estimate the stretch of each explosion model by using the equations in \citet{Guy07} and published $\Delta\textrm{m}_{15}$ values. A brief description of these models and some of their basic physical parameters is presented in Table~\ref{tab:explosion_models}, and more information for each model can be found in the relevant references. This calculation reveals a discrepancy in the fitted relation of Figure \ref{fig:stretch} because, in plotting the predicted mass ratios of each explosion model with respect to specific stretch values, there is no discernible adherence to the trend of other late-time SNe Ia.

While some late-time SNe are visibly closer in stretch and M(${}^{57}$Co)/M(${}^{56}$Co) values to those of explosion models, i.e. SN\,2013aa to the 1.1-0.9 Violent Merger or SN\,2015F to the 0.9-0.76 Violent Merger, there is ultimately no concrete correlation between these specific models and the observed late-time SNe in terms of measured mass ratios and stretch.

To further understand the late-time luminosity evolution of SNe~Ia, we
plot the pseudo-bolometric luminosities of all five SNe~Ia with
extremely late-time data in the right panel of
Figure~\ref{fig:stretch}. Notably, the light curves are very similar
through $\sim$700~days.  After this time, SNe\,2012cg and 2014J have
higher luminosities than that of SNe~2011fe, 2013aa, and 2015F.  In
fact, the latter three SNe have nearly identical light curves (up to
where their data overlap in time) through 1600~days after explosion.
Unsurprisingly, these three SNe have similar isotopic mass ratios, yet there is a noticeable difference between the mass ratio of SN\,2013aa and SN\,2011fe, despite their similar light curve trend. We conclude that the larger mass ratio found in SN\,2011fe is a result of available data in the 500-1000 day phase range in which the relation is presented. The lack of data for SN\,2013aa from 500-1000 days after max light, may be the cause of this lower mass ratio.

SNe\,2012cg and 2014J, on the other hand, have larger
M(${}^{57}$Co)/M(${}^{56}$Co), which has been interpreted as being
the result of having near-Chandrasekhar-mass progenitor stars
\citep{Graur16, yang17}.  However, the measured mass ratios are
significantly larger than that predicted by the SD models.  Moreover,
the difference in mass ratios for SNe\,2012cg and 2014J is larger than
the differences between the different theoretical models.  This
indicates either systematic effects in the luminosity determination
for these SNe, missing physics in the models, or the model parameter
space not spanning the physical parameter space.

\subsection{Non-Optical Contribution to the Bolometric Luminosity} \label{subsec:non-optical}

Since the luminosities calculated for SN\,2013aa are confined to the optical band ($4000 - 9000$\AA), we investigate a non-optical contribution in late-time epochs (particularly at $\sim$10000 -- 20000~\AA). For the case of Fit 1 and 2, the non-optical contribution can be estimated by the ratio of the calculated ${}^{56}\textrm{Co}$ to the total ${}^{56}\textrm{Ni}$, for which we find $\sim$20\% and $\sim$14\%, respectively. These values represent the non-optical $B(t)$ term shown in Section \ref{subsec:bol_lc_model}, and the bolometric correction is found by $1/B(t)$. For the case of Fit 3, in which we fit for this non-optical contribution, we examine the sigmoid function with free parameters generated by the MCMC. We find a gradually increasing non-optical contribution from $\sim$10\% at 100d to $\sim$60\% after 500d from maximum. These values are broadly consistent with the theoretical prediction of $\sim20\%$ by \citet{Fransson15}. Moreover, they are consistent with the non-optical contribution estimations of SN\,2012cg and SN\,2014J based on the \citet{graur17} fits. Based on the infrared analysis of SN~2011fe by \cite{shappee17} and \cite{dimitriadis17}, it is predicted that the non-optical contribution will remain constant after this phase. However, this prediction is based on only one late-time study, and much still remains to be understood about NIR and Mid/Far-Infrared contributions to SN luminosity.

\subsection{Companion Contamination?}

While we see no visible evidence of companion contamination from the photometric analysis or in the \textit{HST} images, we still consider the potential for a surviving binary companion, which could contribute to the luminosity at late-time epochs. We fit the psuedo-bolometric light curve for the decay of ${}^{56}$Co plus a constant companion luminosity.
We calculate a companion contribution to the luminosity of $5.26\pm1.04\times10^{2} \ \textrm{L}_{\odot}$ with a $\chi^2/\textrm{dof} = 297.29/24$. Using the mass-luminosity relation \citep{kuiper38}, this luminosity translates to a main sequence or red giant star with a mass of $\sim 5\:\textrm{M}_{\odot}$. In similar studies such as \citet{dimitriadis17} and \citet{shappee17}, an existing companion star was also ruled out based on the lack of pre- and post-explosion detection. We conclude that this scenario is unlikely in the case of SN\,2013aa.  

\subsection{Light Echoes?}

\citet{graur17} used the late-time SN color evolution to successfully rule out light echo contamination for SN\,2015F, by comparing B-V and V-R with the colors of the well-studied SN\,2011fe, which shows no signs of light echo, having exploded in a relatively clean environment. While we cannot repeat the same procedure for SN\,2013aa, as we do not have this temporal color information at these phase ranges, we can rule out the presence of a light echo by comparing the SED derived from the \textit{HST} $\sim$1500d photometry with SN\,2007af and SN\,2011fe: SN\,2007af is an otherwise normal SN Ia that showed clear signs of a light echo when observed at $\sim$1080d at the same \textit{HST} photometric bands with SN\,2013aa, while for SN\,2011fe, we construct a synthetic SED of the \textit{HST} filters, using the \citet{Taubenberger15} $\sim$1035d spectrum. 

It is straightforward to rule out light echo contamination, as SN\,2013aa is more similar to SN\,2011fe: The SED of SN\,2007af shows the characteristic blue shape of a light echo spectrum, originating from scattered early-time spectra, which is different for both SN\,2013aa and SN\,2011fe. The calculated \textit{F555W}-\textit{F814W} (similar to V-i) colors are 0.79$\pm$0.33, 0.45$\pm$0.02 and -0.49$\pm$0.13 for SN\,2013aa, SN\,2011fe and SN\,2007af, respectively.

\section{Conclusions} \label{sec:conclusion}

In this paper we have presented \textit{HST} WFC3 imaging of SN\,2013aa 1500 days after explosion. Upon detecting the supernova in  three optical filters, we determined the respective AB magnitudes to be 27.969 (\textit{F350LP}), 27.971 (\textit{F555W}), and 27.465 (\textit{F814W}). Based on our astrometric solution, we calculate the chance of coincidence for this detection to be $0.2\%$. Calculated magnitudes at this epoch, combined with photometric data from Swift, LCOGT, and Gemini, allowed for the generation of a pseudo-bolometric light curve. 

In our analysis, we applied the Bateman equation in order to fit the radioactive decays of ${}^{56}\textrm{Ni}$, ${}^{57}\textrm{Ni}$, and ${}^{55}\textrm{Co}$ to the bolometric luminosities of SN\,2013aa. We fit the pseudo-bolometric light curve data with three primary, independent model fits: complete positron trapping (Fit 1), partial positron trapping (Fit 2), and a time-dependent non-optical contribution represented by the sigmoid function (Fit 3). For each model, we estimate the ${}^{57}\textrm{Co}$ and ${}^{55}\textrm{Fe}$ masses and determine the ${}^{57}\textrm{Co}$/${}^{56}\textrm{Co}$ ratio.

For our preferred model (Fit 3), we estimate ${}^{57}\textrm{Co}$/${}^{56}\textrm{Co}=0.02^{+0.01}_{-0.02}$. This value is more consistent with a low-central density, double-degenerate explosion of two sub-Chandrasekhar-mass white dwarf stars than a high-central density Chandrasekhar-mass single-degenerate WD system.

Compared to other SNe~Ia observed at late-time epochs, we find that SN\,2013aa does not match the \citet{graur17} M(${}^{57}$Co)/M(${}^{56}$Co) vs.\ stretch trend. However, the relation presented is for a specific phase range of 500-1000 days, during which SN\,2013aa has no photometric data. Nonetheless, the data at $\sim400$ and $\sim1500$ days is quite constraining in this phase range, and any substantial decrease in luminosity at the 400-500 day or 1000-1500 day phase range is unlikely due to SN\,2013aa's light curve similarity to other late-time SNe Ia. We explore the possibility that the discrepancy in mass ratios may be the result of a major shift in resulting in a substantial non-optical contribution at late-times \citep{Fransson15,sollerman04,leloundas09}. However, if this were the case and SN\,2013aa conformed to the predicted mass ratio by \cite{graur17}, only $\sim 10\%$ of the light at late-times would come from the optical based on our calculated ratio of M(${}^{57}$Co)/M(${}^{56}$Co). While SN\,2013aa may be an outlier to the trend of \cite{graur17}, we cannot sub-classify the target as any different than a normal SN Ia (e.g. 1991T-like) as a result of ambiguity in fitting spectral features.

We note that SN\,2013aa's light-curve evolution and its isotopic mass ratio, are similar to that of SNe\,2011fe. From this similarity in late-time luminosity, we conclude that the slight discrepancy in the masses of SN\,2013aa and SN\,2011fe is the result of missing data between 500-1000 days. Furthermore, we find no direct correlation between the values of stretch and M(${}^{57}$Co)/M(${}^{56}$Co) measured in observed late-time SNe~Ia to those of single-degenerate and double-degenerate explosion models. While the mass ratio and stretch of some late-time SNe Ia are comparable to that of particular explosion models e.g., SN~2013aa to a 1.1+0.9 $\textrm{M}_{\odot}$ Violent Merger, or SN~2015F to a 0.9+0.76 $\textrm{M}_{\odot}$ Violent Merger, there still exists no visible trend between data and models. The significant spread between the predicted mass ratios of explosion models and those of observed late-time SNe indicates a need for either a more comprehensive model analysis of the physics behind supernova explosions, or the reduction of systematic errors in determining the luminosities of late-time SNe~Ia. Additional observations of SN\,2013aa should improve both mass estimates and mitigate potential systematic effects. 

\facilities{HST(WFC3), Gemini:South(GMOS)}

\software{DOLPHOT \citep{dolphot}, emcee \citep{foreman-mackey13}, SNID \citep{Blondin07}, Superfit \citep{Howell05}, IRAF (Tody 1986, Tody 1993), SiFTO \citep{Conley08}, AstroDrizzle \citep{astrodrizzle}}

\begin{acknowledgments} 

We would like to thank David Sand, Dave Coulter, Adam Riess, Dan Scolnic, and Saurabh Jha for helpful comments on this paper. 

Based on observations made with the NASA/ESA {\it Hubble Space
  Telescope}, obtained at the Space Telescope Science Institute
(STScI), which is operated by the Association of Universities for
Research in Astronomy, Inc., under National Aeronautics and Space
Administration (NASA) contract NAS 5--26555. These observations are
associated with Program DD--14925.  Support for
DD--14925 was provided by NASA through a grant from STscI.  This
manuscript is based upon work supported by NASA under Contract No.\
NNG16PJ34C issued through the {\it WFIRST} Science Investigation Teams Program.

The UCSC group is supported in part by NSF grant AST-1518052, the Gordon \& Betty Moore Foundation, and by
fellowships from the Alfred P.\ Sloan Foundation and the David and
Lucile Packard Foundation to R.J.F.\ and from the UCSC Koret Scholars
program to W.V.J.-G.

\end{acknowledgments}

\newpage


\bibliographystyle{aasjournal} 
\bibliography{references}


\end{document}